\documentclass[compsoc, conference, a4paper, 10pt, times]{IEEEtran}

\usepackage{cite}
\usepackage{amsmath,amssymb,amsfonts}
\usepackage{algorithmic}
\usepackage{graphicx}
\usepackage{textcomp}
\usepackage{xcolor}

\def\BibTeX{{\rm B\kern-.05em{\sc i\kern-.025em b}\kern-.08em
    T\kern-.1667em\lower.7ex\hbox{E}\kern-.125emX}}

\usepackage{bbding}
\usepackage{wasysym} 
\usepackage{multirow}
\usepackage{pdflscape} 
\usepackage{csquotes}
\usepackage{color,soul}
\usepackage{hyperref}
\usepackage{listings}

\usepackage{adjustbox}
\usepackage{array}
\newcolumntype{R}[2]{%
    >{\adjustbox{angle=#1,lap=\width-(#2)}\bgroup}%
    c%
    <{\egroup}%
}

\usepackage[inline]{enumitem}

\usepackage{tabularx}
\usepackage{subcaption}
\usepackage{graphicx}

\newcommand{\orcidicon}[1]{\href{https://orcid.org/#1}{\includegraphics[scale=0.2]{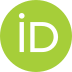}}}
\begin{document}

\title{Evaluation of Provenance Serialisations for Astronomical Provenance}

\author{
    \IEEEauthorblockN{
        Michael A. C. Johnson\IEEEauthorrefmark{1}\IEEEauthorrefmark{2}\IEEEauthorrefmark{3}\orcidicon{0000-0002-5566-6147} \Envelope, Marcus Paradies\IEEEauthorrefmark{1}\IEEEauthorrefmark{4}\orcidicon{0000-0002-5743-6580}, Hans-Rainer Kl\"ockner\IEEEauthorrefmark{2}\orcidicon{0000-0002-0648-2704}, Albina Muzafarova\IEEEauthorrefmark{5}\orcidicon{0000-0002-2282-5105},\\ Kristen Lackeos\IEEEauthorrefmark{2}\orcidicon{0000-0002-6554-3722}, David J. Champion\IEEEauthorrefmark{2}\orcidicon{0000-0003-1361-7723}, Marta Dembska\IEEEauthorrefmark{1}\orcidicon{0000-0002-8180-1525}, Sirko Schindler\IEEEauthorrefmark{1}\orcidicon{0000-0002-0964-4457}
    }
    \IEEEauthorblockA{\IEEEauthorrefmark{1} Institute of Data Science (DLR)}
    \IEEEauthorblockA{\IEEEauthorrefmark{2} Max Planck Institute for Radio Astronomy}
    \IEEEauthorblockA{\IEEEauthorrefmark{3} University of Manchester}
    \IEEEauthorblockA{\IEEEauthorrefmark{4} Technische Universität Ilmenau}
    \IEEEauthorblockA{\IEEEauthorrefmark{5} BETTA Security GmbH}
    \IEEEauthorblockA{\Envelope michael.johnson-4@manchester.ac.uk}
}

\maketitle

\begin{abstract}
Provenance data from astronomical pipelines are instrumental in establishing trust and reproducibility in the data processing and products. 
In addition, astronomers can query their provenance to answer questions routed in areas such as anomaly detection, recommendation, and prediction.
The next generation of astronomical survey telescopes such as the Vera Rubin Observatory or Square Kilometre Array, are capable of producing peta to exabyte scale data, thereby amplifying the importance of even small improvements to the efficiency of provenance storage or querying.
In order to determine how astronomers should store and query their provenance data, this paper reports on a comparison between the turtle and JSON provenance serialisations. 
The triple store Apache Jena Fuseki and the graph database system Neo4j were selected as representative database management systems (DBMS) for turtle and JSON, respectively.
Simulated provenance data was uploaded to and queried over each DBMS and the metrics measured for comparison were the accuracy and timing of the queries as well as the data upload times. 
It was found that both serialisations are competent for this purpose, and both have similar query accuracy.
The turtle provenance was found to be more efficient at storing and uploading the data.
Regarding queries, for small datasets ($<$5MB) and simple information retrieval queries, the turtle serialisation was also found to be more efficient. 
However, queries for JSON serialised provenance were found to be more efficient for more complex queries which involved matching patterns across the DBMS, this effect scaled with the size of the queried provenance. 
\end{abstract}


\section{Introduction}\label{sec:intro}

In recent years, astronomers have exponentially increased their ability to acquire astronomical data.
Final dataset sizes for the next generation of astronomical telescopes is expected to be in the peta or even exa-byte scale.
Datasets of this size require analysis via automated data analysis pipelines.
The users of these pipelines need to rely upon and trust in the results it produces, this reliability and trust can be established by documenting the data processing, i.e. recording provenance. 

In the context of astronomical data, provenance describes the process of generating data.
This includes information such as the original data, the processes applied to it, and the people (or things) responsible for performing those processes. 
PROV is a standard for recording provenance and it models things, processes, and those responsible for processes as entities, activities, and agents, respectively. 
The native format for PROV standard provenance is PROV-N, however it may also be formulated in other formats such as turtle or JSON.
Subsequently, provenance data in these serialisations may then be uploaded to triple stores or graph databases, respectively.
The choice of serialisation therefore dictates both which implementations may be used to store the data as well the language used to query over them.  


This paper therefore provides an evaluation into the efficiency storing and querying astronomical provenance serialised in turtle and JSON.
To perform this evaluation, we simulated a set of provenance data with sizes ranging from $\sim$10KB to $\sim$150MB and serialised the same provenance in each format. 
The turtle data was subsequently uploaded to a triple store, and the JSON data a graph database, using Fuseki and Neo4j as representative implementations. 
These database management systems (DBMS) were then queried over using queried based upon the paper by Johnson et al.~\cite{johnson2021astronomical}, where they outline nine categories of use cases for provenance in astronomy. 
By evaluating each of these use cases within both triple stores and graph databases, we hope to determine which database system has the better query efficiency, storage efficiency, and usability.

The structure of the paper is as follows: Section \ref{sec:app} describes the astronomical applications from which the astronomical provenance was generated; Section \ref{sec:impl} describes how the provenance data was mapped to the rdf and property graph data models and the implementation of our queries; Section \ref{sec:prov} describes the method of provenance generation and simulation; Section \ref{sec:eval} displays the results of the simulations; Section \ref{sec:RW} discusses the related work; Section \ref{sec:conc} presents our conclusions. 

\section{Applications}\label{sec:app}

The base data used for our benchmarking were comprised of provenance that described the function of two astronomical pipelines.
Both pipelines were written in python and they were relatively simple with small ($\sim$10KB) provenance describing their processing.
These base data were then multiplied (via the process described in Section~\ref{sec:prov}) to produce incrementally larger provenance.
The small size of the original provenance data therefore had the advantage of giving greater control over the size of the selected increments and investigating the effectiveness over both small and large provenance datasets. 

\subsection{Optical Imaging - OI}

The purpose of the first astronomical pipeline was to detect populations of stars within optical images.
The star detection was performed using Source Extractor \cite{bertin1996sextractor} on both the original images and those that have been Fourier transformed. 

\subsection{Radio Imaging - RI}

Much like the OI, this pipeline was designed to detect astronomical objects but within radio images.
Consequently, it was composed of different functions, using bdsf\footnote{https://pypi.org/project/bdsf/} for source detection and different methods of transformation prior to plotting.





\section{Queries}\label{sec:impl}

The queries used for this paper were all based upon the use cases outlined in Johnson et al. \cite{johnson2021astronomical}, where the authors outlined nine distinct groups of use cases for astronomical provenance which would be of interest to astronomers.
In order to evaluate these use cases, fourteen pieces of information were identified as important for the evaluation of at least one use case.
Two of these pieces of required information were related to pipeline versions and as all example applications consisted of a single version and underpinning what constitutes a pipeline version is a discussion in and of its own, these requirements were omitted.
For the twelve remaining requirements, a query was written for each that would extract all relevant information that was contained within the provenance.
Table \ref{tab:queries} displays the information that each query was designed to retrieve from the  provenance.
Each query was written in both Cypher and SPARQL and their implementations can be found in Appendix \ref{app:ucs} and can be tested via this "demo-repo"\footnote{\url{https://gitlab.mpcdf.mpg.de/PRAETOR/prov-PRAETOR_public/-/tree/main/prov-PRAETOR/provenance_queries/turtle_vs_rdf}}.

\begin{table}[]\
    \centering
    \begin{tabular}{c|c}
        Query Number & Required Information \\
        \hline
        1 &  Pipeline run identifiers \\
        2 &  Component identifiers \\
        3 &  Data source identifiers \\
        4 &  Data product identifier \\
        5 &  Parameter attributes \\
        6 &  Runtime environment attributes \\
        7 &  Resource consumption \\
        8 &  Data source metadata \\
        9 &  Data product metadata \\
        10 & Quality metrics\\
        11 & Data flow  
    \end{tabular}
    \caption{Desired information to be retrieved by each query.}
    \label{tab:queries}
\end{table}

\section{Provenance}\label{sec:prov}

All provenance was generated with the PRAETOR provenance generation code \cite{praetor},\cite{johnson2024pipeline}.
The PRAETOR code automatically generates PROV compatible provenance for python scripts and the base granularity of which is on the function level. 

Provenance was generated for each of the applications outlined in Section \ref{sec:app}.
Table \ref{tab:filesize} displays the size of the provenance in PROV-N, turtle, and JSON format.
The base format for provenance is PROV-N and the provenance must be reformatted to turtle or JSON in order to be uploaded to the triple store or graph database, respectively. 

\begin{center}
\begin{table}
\begin{tabularx}{\columnwidth}{ | c | c  c  c |}
 Application & PROV-N (KB) & turtle (KB) & JSON (KB) \\ 
 \hline
 Optical Im & 12 & 14 & 22 \\  
 Radio Im & 11 & 13 & 19   
 \end{tabularx}
\caption{Size of original provenance in PROV-N, turtle, and JSON format.}
\label{tab:filesize}
\end{table}
\end{center}

\subsection{Provenance Simulation}

Obtaining enough real data for the large and evenly distributed astronomical provenance datasets needed for our simulations was not practical.
We therefore used the provenance from our applications and multiplied it to simulate the required provenance. 

The process used to generate these data was along the lines of provenance mitosis. The algorithm for which was as follows:

\begin{itemize}
    \item Find all of the identifiers for provenance objects within the base data
    \item Create a new provenance file, where all identifiers for activities and entities have a unique string appended (usually in the form - $s\_n$ where $n$ is the number of times that file has been multiplied already) 
    \item Upload the original and all simulated files to the database
\end{itemize}

\begin{figure*}

\begin{subfigure}{0.48\linewidth}
\centering
  \includegraphics[width=1.05\columnwidth]{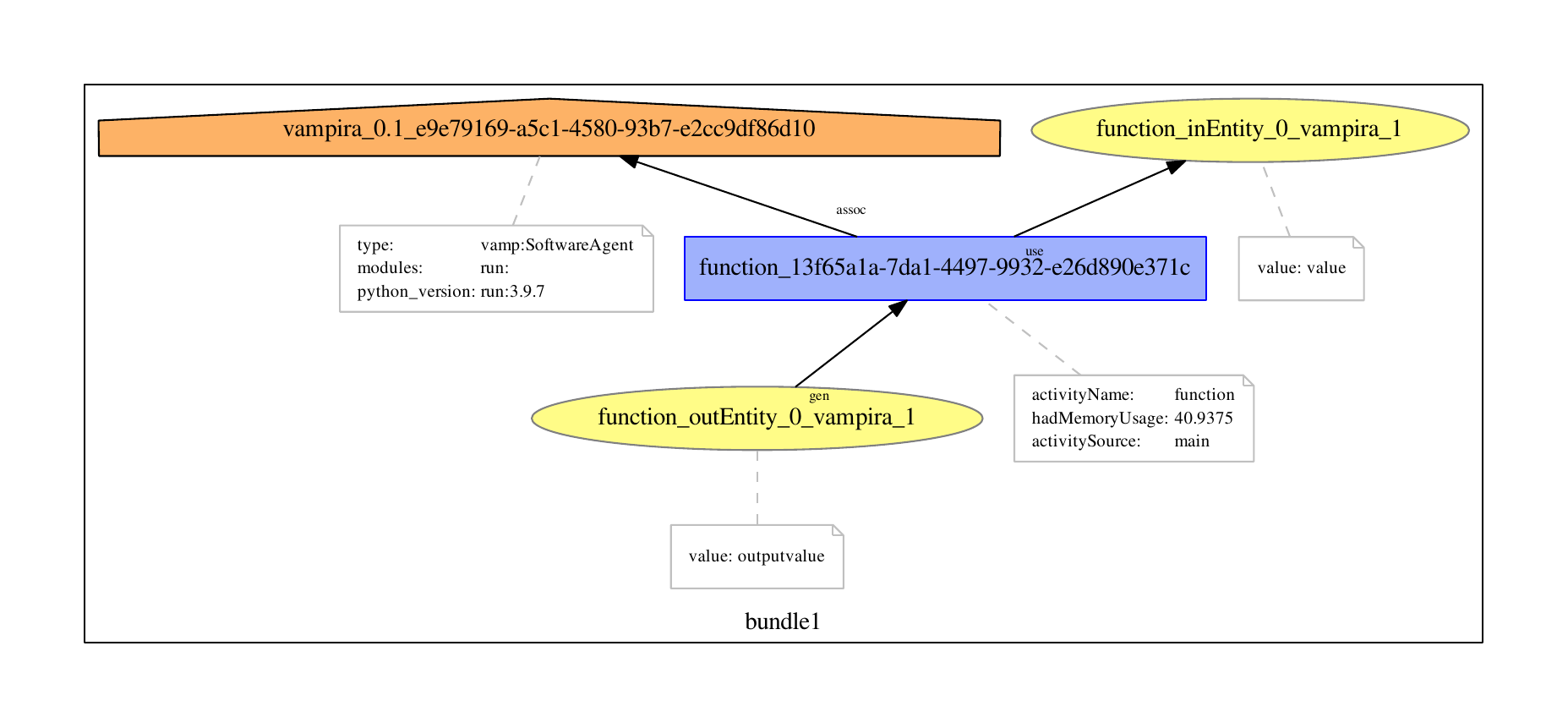}
  \caption{A simple example PROV document depicting an activity which consumes an input entity, produces an output entity, and is attributed to an agent.}
  \label{fig:sim_ex}
\end{subfigure}
\hfill
\begin{subfigure}{0.48\linewidth}
\centering
  \includegraphics[width=\linewidth]{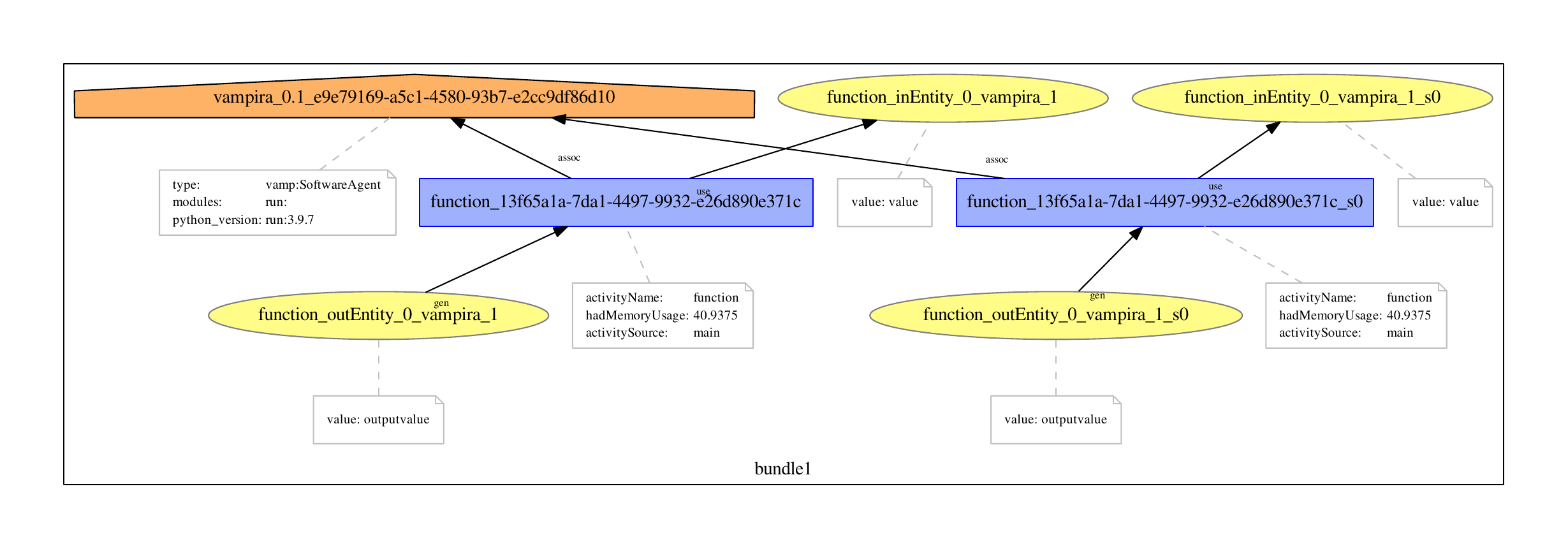}
  \caption{The result of doubling the simple provenance shown in (a), this new graph contains the original activity and entities along with their simulated counterparts, all connected to the original agent. }
  \label{fig:sim_ex2}
\end{subfigure} 
\caption{Provenance mitosis}
\end{figure*}

As only the entities and activities have their identifiers changed, all graphs will still be connected via their agents.
Figure \ref{fig:sim_ex} displays an example of this provenance mitosis in action.

\section{Evaluation}\label{sec:eval}

\subsection{Experimental Setup}

All simulations were run on a Dell Precision 3460, with 16GB of RAM, running Ubuntu 22.04.
All queries for both Neo4j and RDF were formulated in and posted via python. 

Neo4j version 4.1.10 was used throughout the graph database simulations.
The prov2neo python library \cite{de_Boer_prov2neo_2022} was used to import the provenance documents to the Neo4j database.
Cypher was used to query the Neo4j database for all queries to extract the desired information.

The base for the triple store setup was the Apache Jena Fuseki docker image \footnote{https://hub.docker.com/r/secoresearch/fuseki/}.
This docker image was converted to a singularity instance which was used for all triple store simulations.
SPARQL was used to query the Fuseki triple store for all queries to extract the desired information.

\subsection{Query Accuracy}

The quality of the database and query language used to access it were measured using three metrics.
First and foremost, the accuracy of the queries was tested primarily by cross referencing the queries with each other to determine if the results were a match.
If they were, then both queries were assumed to be correct, if not then further investigation would be initiated to determine if either query returned the correct answer. 
As the results from each database would arrive with differing ordering and with different names for the same object, the results were transformed into pandas dataframes and their column names were altered such that they were consistent.
Subsequently, the pandas equals function was leveraged to determine their equivalency. 

\begin{figure}[!htb]

\begin{subfigure}{\linewidth}
  \includegraphics[width=\linewidth]{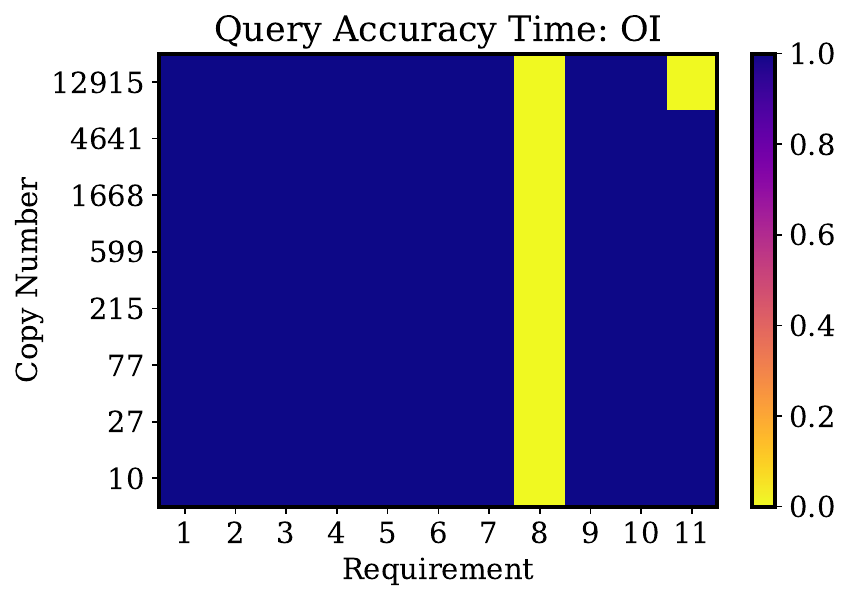}
  \caption{The accuracy of the requirement queries on provenance data from the optical imaging pipeline.}
  \label{fig:eval_acc_opt}
\end{subfigure}\par\medskip

\begin{subfigure}{\linewidth}
  \includegraphics[width=\linewidth]{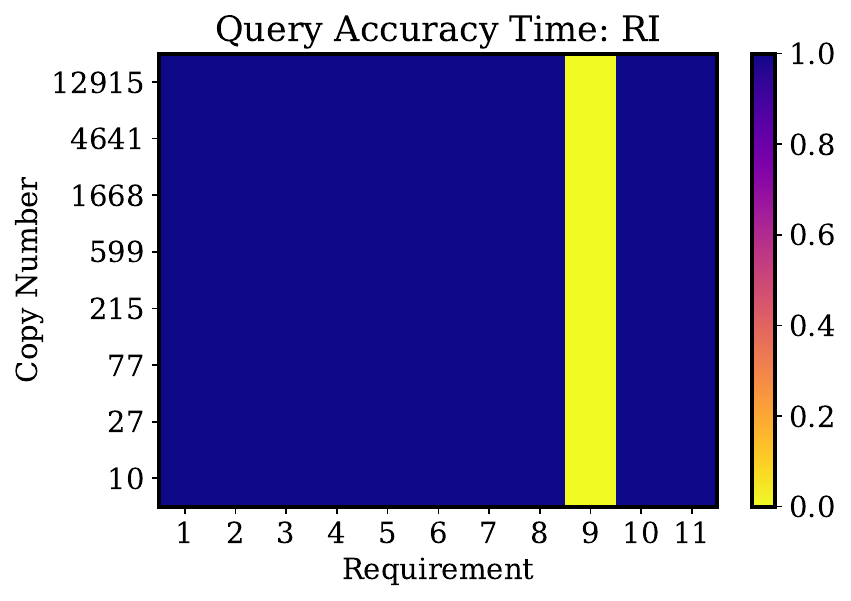}
  \caption{The accuracy of the requirement queries on provenance data from the radio imaging pipeline.}
  \label{fig:eval_acc_rad}
\end{subfigure}

\caption{Query accuracy}
\end{figure}

The results for each of the datasets are shown in Figures \ref{fig:eval_acc_opt}, and \ref{fig:eval_acc_rad}. In each graph, the y axis denotes the number of times the original provenance was multiplied to make the simulated provenance.
The x axis denotes the requirement for which the queries were written.
The colour bar represents whether the output from the Cypher and SPARQL queries matched. 
An accuracy score of 1 is given for a match and zero otherwise.
For the optical imaging pipeline (Figure \ref{fig:eval_acc_opt}) and the radio imaging pipeline (Figure \ref{fig:eval_acc_rad}) the results were consistent across the majority of tested requirements.
The exceptions being requirement 8 for OI and requirement 9 for RI, which were data source metadata and data product metadata respectively.
In each case, there was no data source metadata used in OI or data product metadata used in RI, therefore queries in both Cypher and SPARQL returned blank and these were not counted towards the accuracy. 
The other exception being requirement 11 at the largest simulated provenance dataset size.
Requirement 11 was about data flow and tried to find patterns of connected objects within the queries.
This type of query is shown to be consistently more efficient in Neo4j as shown in Figures \ref{fig:eval_time_opt} and \ref{fig:eval_time_rad}. 
The discrepancy in output from these queries was due to a timeout within the SPARQL query, whereas the Cypher query completed without incident. 
Therefore, the accuracy of the query from SPARQL was reduced, however it was due to the relative inefficiency when compared to its Cypher counterpart.
It should be noted that though the timeout was likely due to a lower efficiency from the RDF implementation, the query would likely execute successfully on a more power machine than the one used within this paper.



\subsection{Query Timing}

\begin{figure}[!htb]
\begin{subfigure}{\linewidth}
  \includegraphics[width=\linewidth]{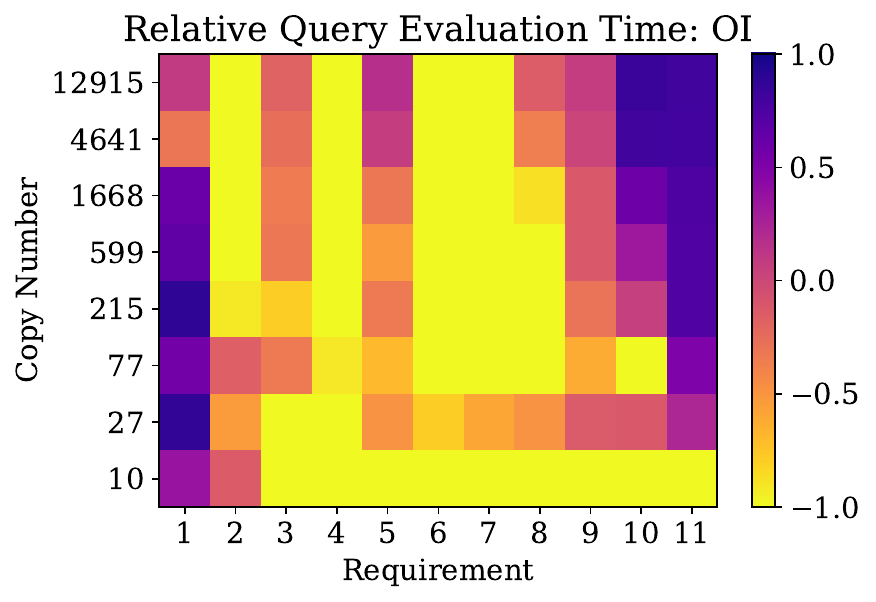}
  \caption{Requirement query execution time comparison for the optical imaging pipeline.}
  \label{fig:eval_time_opt}
\end{subfigure}\par\medskip

\begin{subfigure}{\linewidth}
  \includegraphics[width=\linewidth]{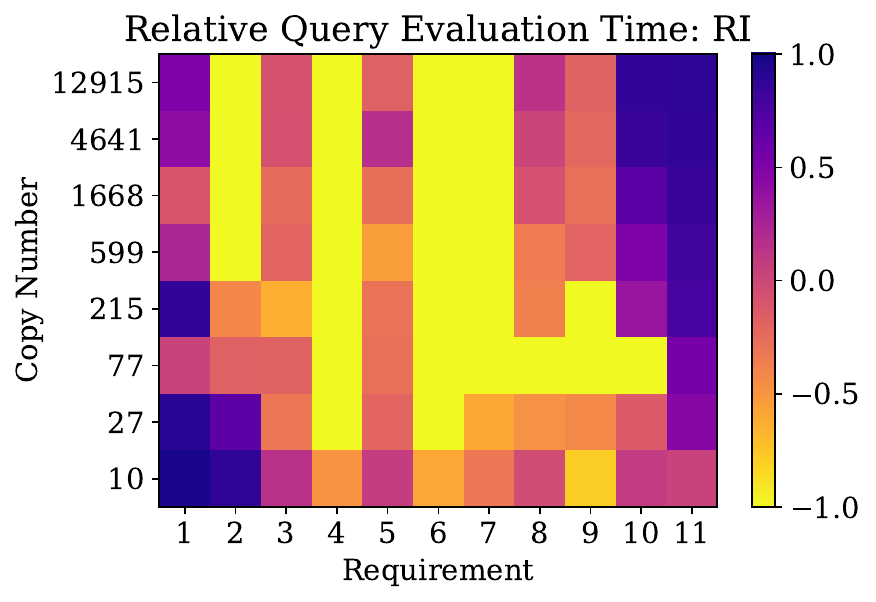}
  \caption{Requirement query execution time comparison for the radio imaging pipeline.}
  \label{fig:eval_time_rad}
\end{subfigure}

\caption{Use case timings}
\end{figure}

The next metric assessed was the timeliness of the queries.
The queries were timed from the start of the request to the database until the results were returned.
Time taken for transformation of the results to dataframes or cross matching were not included.
The results of this timing can be seen in Figures \ref{fig:eval_time_opt} and \ref{fig:eval_time_rad}.
In each graph, the y axis denotes the number of times the original provenance was multiplied to make the simulated provenance.
The x axis denotes the requirement for which the queries were written.
The colour bar in each plot is equal to the $T$ given by Equation \ref{equation} where $n$ and $f$ represent the query execution time in Neo4j and Fuseki, respectively. 
Therefore, yellow areas of the graph represent queries where SPARQL queries were more efficient and the opposite is true for blue regions.
When considering all queries across all dataset sizes, both performed comparably.
However, the Fuseki implementation seems to be more suited for small dataset sizes, whereas Neo4j becomes more efficient the larger the provenance being queried over. 

\begin{equation}\label{equation}
    T = \frac{(f\,-\,n)}{f}
\end{equation}

\subsection{Storage Efficiency}

\begin{figure}[!htb]

\begin{subfigure}{\linewidth}
  \includegraphics[width=\linewidth]{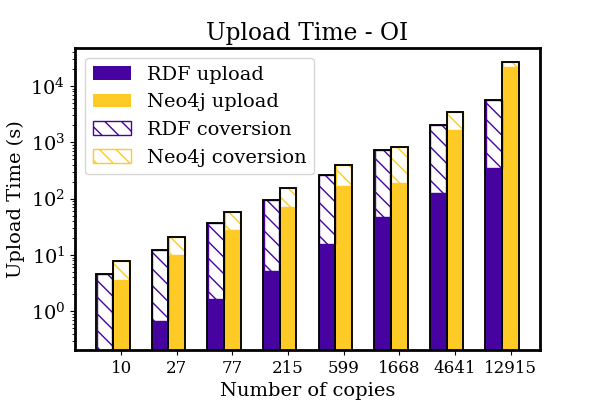}
  \caption{The upload and conversion time for provenance simulated from the optical imaging pipeline.}
  \label{fig:up_time_opt}
\end{subfigure} \par\medskip

\begin{subfigure}{\linewidth}
  \includegraphics[width=\linewidth]{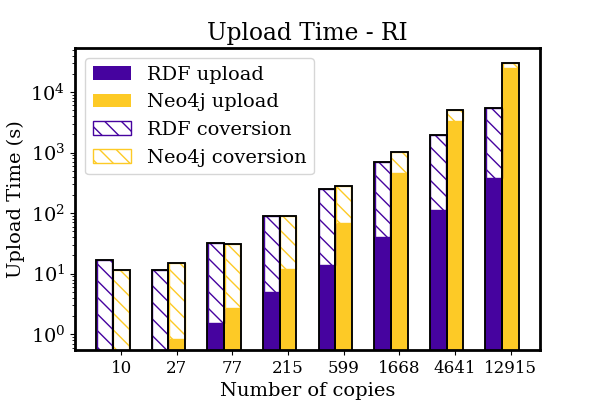}
  \caption{The upload and conversion time for provenance simulated from the radio imaging pipeline.}
  \label{fig:up_time_rad}
\end{subfigure} 

\caption{Upload timings}
\end{figure}

The efficiency of storage was also considered in two separate forms, firstly the efficiency of storing provenance information in the file format required for each database, shown previously in Table \ref{tab:filesize}.
Secondly, the time taken to convert and upload the PROV-N format provenance to each database, shown in Figures \ref{fig:up_time_opt} and \ref{fig:up_time_rad} for the OI and RI pipelines, respectively. 
For each graph, the x axis denotes the number of times the original provenance was multiplied to make the simulated data and the y axis denotes the time take to upload the data.
The solid bar represents the upload time whilst the stacked, hashed bar represent the time needed for conversion.
The conversion for the RDF implementation consisted of using the ProvToolBox to convert the provenance from PROV-N format to turtle.
The conversion for the Neo4j implementation also used the ProvToolBox to convert from PROV-N to JSON but then also utilised the prov2neo python package \cite{de_Boer_prov2neo_2022} to make the provenance compatible for upload.

For all provenance dataset size, upload times were lower in the RDF implementation when compared to the Neo4j counterpart.
However, times for the combination of upload and conversion were similar between each system for dataset sizes up to 215 copies or roughly 2.5MB of PROV-N data.
For provenance datasets beyond this size, the RDF implementation was faster when considering both upload and conversion times.
In order for all of the data to fit within one graph in Figures \ref{fig:up_time_opt} and \ref{fig:up_time_rad}, a log time scale was used. 
Therefore, whilst the bar chart effectively displays the comparison between the comparison between Neo4j and RDF upload time and combined upload and conversion times, it can be misleading when comparing upload and conversion time within a single implementation or conversion times between implementations.
In order to enable comparison between conversion times, Appendix \ref{app:timing} contains Table \ref{tab:upconv} which displays the upload and conversion times within each implementation for each tested size of provenance for the OI and RI pipelines. 

\section{Related Work}\label{sec:RW}

The related work is divided into two sections.
The first is dedicated to benchmarking various database systems.
The majority of the papers in this section use non-provenance data for their benchmarking simulations.
However, Vicknair et. al. \cite{vicknair2010comparison} and Glavic et. al. \cite{glavic2014big} explicitly use provenance data.
The second section is devoted to examples of provenance querying in papers, using a single database system for all queries. 

\subsection{Benchmarking Database Systems}

In Vicknair et. al. \cite{vicknair2010comparison}, they explore the possibility of storing provenance data in NoSQL databases, as apposed to the traditional SQL implementations.
Their database systems of choice were Neo4j and MySQL for the NoSQL and SQL databases, respectively.
They constructed twelve databases within each system which were each composed of random graphs.
They then constructed three queries to extract structural information and three to extract data to be evaluated on each database.
They found that Neo4j outperformed MySQL on all structural queries, however for their randomly generated data, MySQL outperformed Neo4j on data type queries.
After updating their data to more resemble real data, it was found that Neo4j also outperforms MySQL on data type queries but only on large scale databases.
Within their conclusions, they noted that Neo4j can be converted to and queried as an RDF store which may have the potential to alleviate some of their tertiary concerns surrounding the Neo4j implementation.

In Glavic et. al. \cite{glavic2014big}, the importance of provenance for big data systems is discussed as well as the potential challenges for the benchmarking of it. 
Scalability and robustness are identified as the prime measures for provenance benchmarking. 
A number of generic use cases are also discussed to exemplify the usefulness of provenance in big data such as the identification of poorly performing pipeline components.

In Angles et. al. \cite{angles2013benchmarking}, they set out to benchmark a selection of database systems for their handling and querying of social network data.
Their chosen database systems were: Neo4j and Dex to represent graph databases; RDF-3X for resource description frameworks; Virtuoso and PostgreSQL for relational databases.
They generated twelve queries to extract relevant information and benchmarked their  database systems using the following metrics: data loading time, query execution time, and data indexes.
They found that in general, all database systems performed well. However, they found that the graph databases generally performed the best and that relational databases struggled to complete some of the proposed queries in a reasonable time frame. 

In their paper Pobiedina et. al. \cite{pobiedina2014benchmarking} benchmark a variety of database systems specifically for their competency with pattern matching.
Their chosen database systems are: PostgreSQL, RDF (Jena), Neo4j, and Clingo.
They populated each system with a combination of simulated and real data, repeating patterns were added within the simulated data and were inherent to the real data.
After searching for these patterns within their data, they found that the RDF and graph database systems generally performed the best, with Neo4j struggling to find cyclic patterns and RDF performing the best overall.
However, they note that at the time of writing, none of their systems were sufficiently developed to handle pattern matching on a large scale in real data. 

In their paper Dominguez et. al. \cite{dominguez2010survey} benchmarked the database systems Neo4j, Jena, HypergraphDB, and DEX.
The queries used were those of the HPC Scalable Graph Analysis Benchmark v1.0. They found that for small dataset sizes, all database systems performed well.
However, only DEX and Neo4j were capable of loading the largest benchmark sizes.

\subsection{Querying Provenance}

In their paper Chebotko et. al. \cite{chebotko2007storing} investigated storing and querying provenance via a relational database management system, translating it from RDF. 
They mapped their RDF data to relational format and created an algorithm to transform SPARQL queries to SQL.
They found that their translations were efficient and scalable whilst also offering the storage and querying power expected of a RDMS.

In their paper Johnson et. al. \cite{johnson2018using} used SPARQL queries over provenance that described the operation of astronomical pipelines.
The queries were used to evaluate a number of different use cases that would be important to astronomers and demonstrated the advantages that recording and querying provenance can offer.

\section{Discussion and Conclusions}\label{sec:conc}

In order to inform the choice of data model for provenance, the graph database system Neo4j and the triple store Fuseki were evaluated to determine each DBMSs competency for storing and querying provenance data from astronomical pipelines.
The metrics used for evaluation were the upload time, storage efficiency, query accuracy, and query timing.
It was found that each system tested performed comparatively well in each category.

At small ($\leq$5MB) provenance data sizes, the combined time for upload and conversion of provenance was comparable for both DBMSs.
However, at higher dataset sizes the Fuseki triple store marginally outperformed Neo4j is this category.

Both pipelines return the same results for most requirement queries. 
Discrepancies were observed in returned results from requirements 7 and 8 in the OI and RI pipeline respectively.
Both of which returned null results as these pipelines did not contain the information those queries were searching for.
The only other discrepancy found was in the largest dataset for requirement 11 where the SPARQL query timed out due to its relatively inefficiency at pattern queries in large datasets when compared to Cypher.

Query times between the two implementations were dependant on the type of query performed. 
Queries 1, 2, 6, and 7 were all simple information queries which required no pattern. 
Each of these queries was faster in SPARQL rather than Cypher with the exception of 1.
The unique quality of query 1 is that it returns information on an agent, of which there is only one per provenance graph.
The only other query where the SPARQL query was faster than the Cypher was query 4 - a {\it very} simple pattern match query. 
Every other query displayed the following trend, the larger the dataset, the more likely the Neo4j implementation had the faster query.
The size of the dataset was not the only trend displayed in Figures \ref{fig:eval_time_opt} and \ref{fig:eval_time_rad}, it also displayed that with more complex queries, they would become faster in Cypher than SPARQL at smaller dataset sizes. 

In conclusion, it was found that the RDF implementation more efficiently stored the tested provenance. 
The turtle format was also more efficient provenance serialisation. 
Finally, the SPARQL queries within the chosen RDF implementation were faster than that of Neo4j and Cypher when the queries were very simple, with little to no pattern matching. 
Conversely, with more complicated queries the Cypher queries within Neo4j were faster than that of SPARQL, particularly at larger provenance dataset sizes. 
Therefore, the choice of provenance data serialisation for astronomical provenance should be motivated the manner in which it will be used - for simple information retrieval use turtle and for more complex use cases JSON.

\section*{Acknowledgement}

Supported by the German Federal Ministry for Economic Affairs and Energy on the basis of a decision of the German Bundestag under the project number 50OO1905.
This work is supported by the UK SKA Regional Centre which receives funding from UKRI-STFC.

{\footnotesize \bibliographystyle{acm}
\bibliography{sample}}

\appendix



\section{Use case Queries}\label{app:ucs}
In the examples we made use of the following prefixes:
\begin{itemize}
    \item \texttt{prov} \ldots \url{http://www.w3.org/ns/prov#}
    \item \texttt{run} \ldots \url{http://example.org/}
    \item \texttt{prtr} \ldots \url{https://praetor.pages.mpcdf.de/prov-PRAETOR_public/}
    \item \texttt{urn\_uuid} \ldots \url{urn:uuid:}
\end{itemize}

\subsection*{Requirement 1 - Pipeline Run Identifier}

\subsubsection*{Cypher}

\begin{lstlisting}[linewidth=\columnwidth,breaklines=true]
MATCH (n:Agent) 
RETURN n.`prov2neo:identifier`
\end{lstlisting}
\subsubsection*{SPARQL}

\begin{lstlisting}[linewidth=\columnwidth,breaklines=true]
PREFIX prov: <http://www.w3.org/ns/prov#>
SELECT ?pid
FROM NAMED <pipeline>
WHERE {
GRAPH ?g {
?pid a prov:Agent .
}
}
\end{lstlisting}

\subsection*{Requirement 2 - Component Identifiers}

\subsubsection*{Cypher}

\begin{lstlisting}[linewidth=\columnwidth,breaklines=true]
MATCH (n:Activity) 
RETURN n.`prov2neo:identifier`
\end{lstlisting}

\subsubsection*{SPARQL}

\begin{lstlisting}[linewidth=\columnwidth,breaklines=true]
PREFIX prov: <http://www.w3.org/ns/prov#>
SELECT ?cid
FROM NAMED <pipeline>
WHERE {
GRAPH ?g {
?cid a prov:Activity .
}
}
\end{lstlisting}

\subsection*{Requirement 3 - Data Source Identifier}

\subsubsection*{Cypher}

\begin{lstlisting}[linewidth=\columnwidth,breaklines=true]
MATCH p=()-[r:used]->(n) 
RETURN n.`prov2neo:identifier`
\end{lstlisting}

\subsubsection*{SPARQL}

\begin{lstlisting}[linewidth=\columnwidth,breaklines=true]
PREFIX prov: <http://www.w3.org/ns/prov#>
SELECT ?dsid
FROM NAMED <pipeline>
WHERE {
GRAPH ?g {
?b a prov:Usage ;
    prov:entity ?dsid .
}
}
\end{lstlisting}

\subsection*{Requirement 4 - Data Product Identifier}

\subsubsection*{Cypher}

\begin{lstlisting}[linewidth=\columnwidth,breaklines=true]
MATCH p=(n)-[r:wasGeneratedBy]->() 
WHERE EXISTS(n.`prov2neo:identifier`) 
RETURN n.`prov2neo:identifier`
\end{lstlisting}

\subsubsection*{SPARQL}

\begin{lstlisting}[linewidth=\columnwidth,breaklines=true]
PREFIX prov: <http://www.w3.org/ns/prov#>
SELECT ?dpid
FROM NAMED <pipeline >
WHERE {
GRAPH ?g {
?dpid prov:wasGeneratedBy ?a
}
}
\end{lstlisting}
\subsection*{Requirement 5 - Parameter Attributes}

\subsubsection*{Cypher}

\begin{lstlisting}[linewidth=\columnwidth,breaklines=true]
MATCH p=()-[r:used]->(n) 
WHERE EXISTS(n.`prov:value`) 
RETURN n.`prov:value`
\end{lstlisting}

\subsubsection*{SPARQL}

\begin{lstlisting}[linewidth=\columnwidth,breaklines=true]
PREFIX prov: <http://www.w3.org/ns/prov#>
SELECT ?pi
FROM NAMED <pipeline>
WHERE {
GRAPH ?g {
?b a prov:Usage ;
    prov:entity ?dsid .
?dsid prov:value ?pi .
}
}
\end{lstlisting}

\subsection*{Requirement 6 - Runtime Environment Attributes}

\subsubsection*{Cypher}

\begin{lstlisting}[linewidth=\columnwidth,breaklines=true]
MATCH (n) 
WHERE EXISTS (n.`prtr:modules`) 
AND EXISTS (n.`prtr:python_version`) 
RETURN n.`prtr:modules`, n.`prtr:python_version`
\end{lstlisting}

\subsubsection*{SPARQL}

\begin{lstlisting}[linewidth=\columnwidth,breaklines=true]
PREFIX prov: <http://www.w3.org/ns/prov#>
PREFIX prtr: <https://praetor.pages.mpcdf.de/prov-PRAETOR_public/> 
SELECT ?m ?pv
FROM NAMED <pipeline>
WHERE {
GRAPH ?g {
?a a prov:Agent ;
    prtr:modules ?m ;
    prtr:python_version ?pv .
}
}
\end{lstlisting}

\subsection*{Requirement 7 - Resource Consumption Attributes}

\subsubsection*{Cypher}

\begin{lstlisting}[linewidth=\columnwidth,breaklines=true]
MATCH (n)
WHERE EXISTS(n.`prov:startTime`) 
AND EXISTS(n.`prov:endTime`) 
AND EXISTS(n.`prtr:hadMemoryUsage`)
RETURN n.`prov:startTime`, n.`prov:endTime`, n.`prtr:hadMemoryUsage`
\end{lstlisting}

\subsubsection*{SPARQL}

\begin{lstlisting}[linewidth=\columnwidth,breaklines=true]
PREFIX prov: <http://www.w3.org/ns/prov#>
PREFIX prtr: <https://praetor.pages.mpcdf.de/prov-PRAETOR_public/> 
SELECT ?st ?et ?m
FROM NAMED <pipeline>
WHERE {
GRAPH ?g {
?a a prov:Activity ;
    prov:startedAtTime ?st ;
    prov:endedAtTime ?et ;
    prtr:hadMemoryUsage ?m .
    }
    }
\end{lstlisting}

\subsection*{Requirement 8 - Data Source Metadata Attributes}

\subsubsection*{Cypher}

\begin{lstlisting}[linewidth=\columnwidth,breaklines=true]
MATCH p=(n)-[r]-(m)
WHERE EXISTS (r.`prov:role`) 
AND EXISTS (m.`prov:value`) 
AND r.`prov:role` = "run:fileAccess_mode" 
AND m.`prov:value` = "r" 
WITH n MATCH q=(n)-[r:used]-(e) 
RETURN e.`prov:value`
\end{lstlisting}

\subsubsection*{SPARQL}

\begin{lstlisting}[linewidth=\columnwidth,breaklines=true]
PREFIX prov: <http://www.w3.org/ns/prov#>
PREFIX run: <http://example.org/>
PREFIX prtr: <https://praetor.pages.mpcdf.de/prov-PRAETOR_public/> 
SELECT ?evv 
FROM NAMED <pipeline>
WHERE {
GRAPH ?g {
?a prtr:activityName "pythonBuiltinFileAccess" ;
    prov:qualifiedUsage ?b .
?b prov:entity ?eid ;
        prov:hadRole run:fileAccess_mode .
?eid prov:value ?ev .
?a prov:qualifiedUsage ?b2 .
?b2 prov:entity ?e2 .
?e2 prov:value ?evv .
filter(?ev = "r")
}
}
\end{lstlisting}

\subsection*{Requirement 9 - Data Product Metadata Attributes}

\subsubsection*{Cypher}

\begin{lstlisting}[linewidth=\columnwidth,breaklines=true]
MATCH p=(n)-[r]-(m)
WHERE EXISTS (r.`prov:role`)
AND EXISTS (m.`prov:value`)
AND r.`prov:role` = "run:fileAccess_mode"
AND m.`prov:value` = "w"
WITH n MATCH q=(n)-[r:used]-(e) 
RETURN e.`prov:value`
\end{lstlisting}

\subsubsection*{SPARQL}

\begin{lstlisting}[linewidth=\columnwidth,breaklines=true]
PREFIX prov: <http://www.w3.org/ns/prov#>
PREFIX run: <http://example.org/>
PREFIX prtr: <https://praetor.pages.mpcdf.de/prov-PRAETOR_public/> 
SELECT ?evv 
FROM NAMED <pipeline>
WHERE {
GRAPH ?g {
?a prtr:activityName "pythonBuiltinFileAccess" ;
    prov:qualifiedUsage ?b .
?b prov:entity ?eid ;
    prov:hadRole run:fileAccess_mode .
?eid prov:value ?ev .
?a prov:qualifiedUsage ?b2 .
?b2 prov:entity ?e2 .
?e2 prov:value ?evv .
filter(?ev = "w")
}
}
\end{lstlisting}

\subsection*{Requirement 10 - Quality Metric Attributes}

\subsubsection*{Cypher}

\begin{lstlisting}[linewidth=\columnwidth,breaklines=true]
MATCH (n) WHERE EXISTS(n.`rdf:type`) RETURN DISTINCT n.`prov2neo:identifier`
\end{lstlisting}

\subsubsection*{SPARQL}

\begin{lstlisting}[linewidth=\columnwidth,breaklines=true]
PREFIX prov: <http://www.w3.org/ns/prov#>
PREFIX prtr: <https://praetor.pages.mpcdf.de/prov-PRAETOR_public/> 
SELECT ?q
FROM NAMED <pipeline>
WHERE {
GRAPH ?g {
?a prtr:hadQuality ?q .
?q prov:value ?v ;
    a ?qt .
FILTER(?qt != prov:Entity) 
}
}
\end{lstlisting}

\subsection*{Requirement 11 - Data Flow Connections}

\subsubsection*{Cypher}

\begin{lstlisting}[linewidth=\columnwidth,breaklines=true]
MATCH p=(e1)-[r1:wasGeneratedBy]->(a1)-[r2:used]->(e2)-[r3:wasGeneratedBy]->(a2)-[r4:used]->(e3) 
RETURN e1.`prov2neo:identifier`, a1.`prov2neo:identifier`, e2.`prov2neo:identifier`, a2.`prov2neo:identifier`, e3.`prov2neo:identifier`
\end{lstlisting}

\subsubsection*{SPARQL}

\begin{lstlisting}[linewidth=\columnwidth,breaklines=true]
PREFIX prov: <http://www.w3.org/ns/prov#>
SELECT ?e ?a ?e2 ?a2 ?e3
FROM NAMED <pipeline>
WHERE {
GRAPH ?g {
?a a prov:Activity ;
    prov:qualifiedUsage ?b .
?b prov:entity ?e .
?e2 prov:wasGeneratedBy ?a .
?a2 a prov:Activity ;
    prov:qualifiedUsage ?b2 .
?b2 prov:entity ?e2 .
?e3 prov:wasGeneratedBy ?a2
}
}
'''
\end{lstlisting}

\section{Upload and Conversion Times}\label{app:timing}

\begin{adjustbox}{angle=90, caption={Upload and conversion times within RDF and Neo4j for the OI and RI pipelines.}, label={tab:upconv}, float=table}   

\centering
    \begin{tabular}{c||c | c | c | c | c | c | c | c}
        No. of copies & 10 & 27 & 77 & 215 & 599 & 1668 & 4641 & 12915 \\
         \hline
        OI Pipeline & & & & & & & &  \\
        \hline
        Neo4j Upload Time (s) &  0.4 & 0.9 & 3.3 & 13.5 & 73.0 & 478.3 & 3512.0 & 26528.6\\
        Neo4j Conversion Time (s) &  4.1 & 10.5 & 30.3 & 81.8 & 229.1 & 638.5 & 1786.0 & 4939.6 \\
         \hline
        RDF Upload Time (s) &  0.2 & 0.7 & 1.6 & 5.2 & 15.9 & 46.5 & 125.3 & 355.2 \\
        RDF Conversion Time (s) & 4.4 & 11.6 & 35.0 & 88.2 & 243.3 & 686.2 & 1903.0 & 5272.6 \\
        \hline
        RI Pipeline & & & & & & & &  \\
        \hline
        Neo4j Upload Time (s) &  0.3 & 0.8 & 2.7 & 12.0 & 68.1 & 451.7 & 3330.5 & 25196.9\\
        Neo4j Conversion Time (s) &  11.3 & 14.1 & 28.0 & 77.1 & 212.2 & 588.8 & 1644.9 & 4600.7 \\
         \hline
        RDF Upload Time (s) &  0.4 & 0.5 & 1.5 & 4.9 & 13.8 & 40.5 & 113.5 & 374.7 \\
        RDF Conversion Time (s) & 16.4 & 10.8 & 30.5 & 85.1 & 234.5 & 650.0 & 1812.6 & 5088.5 \\
    \end{tabular}
\end{adjustbox}


\end{document}